\documentclass[conference]{IEEEtran}
\IEEEoverridecommandlockouts
\usepackage{cite}
\usepackage[hidelinks,bookmarks=false]{hyperref}
\usepackage{amsmath,amssymb,amsfonts}
\usepackage{dblfloatfix} 
\usepackage{algorithm}
\usepackage{algpseudocode}
\usepackage{graphicx}
\usepackage{textcomp}
\usepackage{listings}
\usepackage[table]{xcolor}
\usepackage{tikz}
\usetikzlibrary{calc}
\usepackage{nomencl}

\usepackage{float}
\usetikzlibrary{shadows.blur}   
\usepackage{circuitikz}

\tikzset{%
        brace/.style = { decorate, decoration={brace, amplitude=5pt} },
       mbrace/.style = { decorate, decoration={brace, amplitude=5pt, mirror} },
        label/.style = { black, midway, scale=0.5, align=center },
     toplabel/.style = { label, above=.5em, anchor=south },
    leftlabel/.style = { label,rotate=-90,left=.5em,anchor=north },   
  bottomlabel/.style = { label, below=.5em, anchor=north },
        force/.style = { rotate=-90,scale=0.4 },
        round/.style = { rounded corners=2mm },
       legend/.style = { right,scale=0.4 },
        nosep/.style = { inner sep=0pt },
   generation/.style = { anchor=base }
}

\hypersetup{
	colorlinks=true, 
	linkcolor=blue, 
	citecolor=blue, 
	urlcolor=green 
}

\begin{document}

\title{FractalSortCPU: Bandwidth-Efficient Compressed Radix Sort on CPU}

\author{\IEEEauthorblockN{Michael Dang'ana}
\IEEEauthorblockA{\textit{Electrical \& Computer Engineering} \\
\textit{University of Toronto}\\
\textit{Eonforge Labs}\\
Toronto, Canada \\
michael.dangana@mail.utoronto.ca}
}

\maketitle

\begin{abstract}
Cloud database systems, particularly their middleware and query execution layers, use sorting as a core operation in query processing, indexing and join execution. Distribution-dependence and limited parallelism are key issues inherent in state-of-the-art radix sort which is preferred for large datasets due to performance advantages over comparison-based algorithms. Multi-pass bucketing, stochastic sampling and dependence graph structures are common solutions to these problems that incur the cost of data pre-processing and increased memory footprint hence they are less appropriate for large-scale workloads common in cloud environments. In-place radix sort schemes increase the number of passes as precision increases, which negatively impacts latency. \\

Our work solves these problems by introducing a CPU-adapted histogram compression scheme for radix sorting for arbitrary-precision keys implemented on the CPU for increased accessibility, providing state-of-the-art execution time, while limiting histogram growth. Fully parallel key-based histogram updates eliminate the need for input bucketing and data pre-processing further lowering latency, mitigating distribution-dependence and reducing complexity. With a parallelized sorting architecture utilizing SIMD-accelerated operations for low latency, the algorithm demonstrates improvement over the state-of-the-art on the CPU, GPU, and FPGA by 6x, 3x and 2.5x in bandwidth efficiency on 512MB to 32GB data sets at 16-bit precision. 

\end{abstract}

\section{Introduction}
As workloads grow and query demand tightens, cloud database middleware and query execution systems face large-scale sorting challenges. Sorting is a key concern in big data scenarios and is impacted by growing datasets and resource scarcity, increasing the latency, throughput and consistency requirements locally and in the cloud \cite{MLGPUShortage} \cite{KsurfDrone}. Sorting enables big data use cases for operations such as index creation, sort-merge joins, queries, and database partitioning \cite{sortingUsesDb}. Radix sort algorithm is ideal for larger problem sizes compared to fast approaches based on sorting networks, merge sort, and sample sort \cite{CaseStudyRadixSort} \cite{satish2009designing} \cite{satish2010fast}. \\

State-of-the-art radix sort implementations usually suffer from \textit{distribution dependence} where skewed distributions lead to imbalanced bucket sizes affecting efficiency due to the need to suffer degradation from larger buckets or to dynamically allocate processing resources to large buckets, \textit{limited parallelism} due to multiple passes required for bucketing, \textit{linear precision-scaling latency} where extra passes are necessary to achieve high precision sorts, and \textit{memory usage growth} where pre-processing techniques including stochastic sampling, distribution histograms and dependency graphs require data structures impacting memory usage \cite{Bingmann2018} \cite{Blelloch1991} \cite{satish2009designing} \cite{Sanders1997} \cite{Dehne1996} \cite{Ajwani2016}. \\

Distribution dependence refers to the varying performance of sorting algorithms based on the inputs  regarding skew, partially sorted data, randomness, uniform and Gaussian distribution from a latency and memory usage perspective \cite{DistribDepInplaceSorting}. Sorting algorithms are usually evaluated in terms of best, average and worst case performance which are significantly impacted by input distribution \cite{DistribDepPerfMetricsDistribDepli2025}. Various techniques are used to mitigate pre-processing necessary to determine input distribution such as adaptive  partitioning and sorting strategy adjustment enabling consistent performance across a range of input distributions \cite{DistribDepAdaptivePartitioning2025optiflexsort}. \\

Limited parallelism occurs where high-performance radix sort uses multiple passes to partially sort at lower precision and store the intermediate data in memory, with a significant probability of histogram bucket collisions between entries at each pass creating critical paths \cite{Ajwani2016}. Platform bottlenecks such as limited cache and socket capacity on CPU, memory conflicts and warp divergence cause bottlenecks and core idling on GPU further limiting effective parallelism, mitigated by techniques including bucket balancing and increased memory bandwidth \cite{WarpDivergenceAnantpur2017}. \\

The precision-scaling latency trade-off is characterized by the increase in sorting latency with precision where reduced precision to achieve approximate sorting is significantly beneficial in terms of reducing latency by over $40\%$ using approximate storage, and full precision can be achieved by iteratively sorting the partially sorted data, relevant in modern hardware algorithms where reductions in compute and memory access can lead to significant performance gains for large data sets \cite{PrecisionLatencyTradeoffShuang2016}. \\

Memory utilization in radix distribution histograms can grow exponentially with precision necessitating multiple passes to accomodate memory capacity and memory access bandwidth limits on GPUs, FPGAs and CPUs for full precision sorting \cite{hrs}. Temporary buffers for histogram population, sampling and bucket balancing further increase memory demand, mitigated by techniques such as adjusting the number of keys per pass to enable near-linear precision-scaling at the cost of increased memory, cache and bandwidth usage \cite{Ajwani2016} \cite{Inoue2007}. \\

This work extends FractalSort \cite{FractalSortFPGA}, originally designed for FPGA, to a CPU-adapted implementation that introduces the concept of the fractal data structure, a sparse binary tree histogram. This enables sublinear execution time for large datasets, SIMD-enabled thread execution of histogram updates, independent key processing without input bucketing or partitioning, and amortized memory overhead for large datasets due to histogram compression and localized histogram updates. The contributions include:

\begin{enumerate}
    \item A parallel sort and merge update for unsorted keys with amortized update contention.
    \item A novel batch stream sorting algorithm for large datasets without data pre-processing or input bucketing.
    \item A novel sparse histogram compression scheme for bandwidth-efficient merge operation with counter width tapering.
    \item A novel bandwidth-efficient FractalSortCPUA algorithm to retrieve an array of sorted keys from the compressed histogram. \\
\end{enumerate}  

Addressing these challenges enables FractalSortCPU to offer an accessible, scalable solution for cloud servers, databases and distributed big data compute systems while lowering costs and query latency at scale \cite{FractalSortCPUCode}, complementing cloud resource optimization approaches \cite{KsurfDrone}. \\

\section{Related Work}
Radix sorting is the state-of-the-art algorithm for large keys due to the linear complexity with implementations on both serial and parallel architectures, including least significant bit (LSB) and most significant bit (MSB) algorithms often divided into histogram, prefix sum and stable scatter phases \cite{Knuth1998} \cite{CaseStudyRadixSort}. One way to determine bucket bounds is coarse-grained bucketing combined with prefix scanning that offer speed-ups on uniform distributions but suffer from distribution dependence \cite{Hagerup1991} \cite{Leischner2010}. Unit processor local histograms with reduction, scanning and merge scatter phases are common on GPU, memory and cache-locality are prevalent on CPU, while streaming data paths are common on FPGA due to higher memory access speeds, with memory bandwidth, radix size and precision-scalability the primary design concerns \cite{hrs} \cite{Inoue2007}.  \\

\begin{figure*}[!htbp]
    \centering
    \includegraphics[scale=0.485]{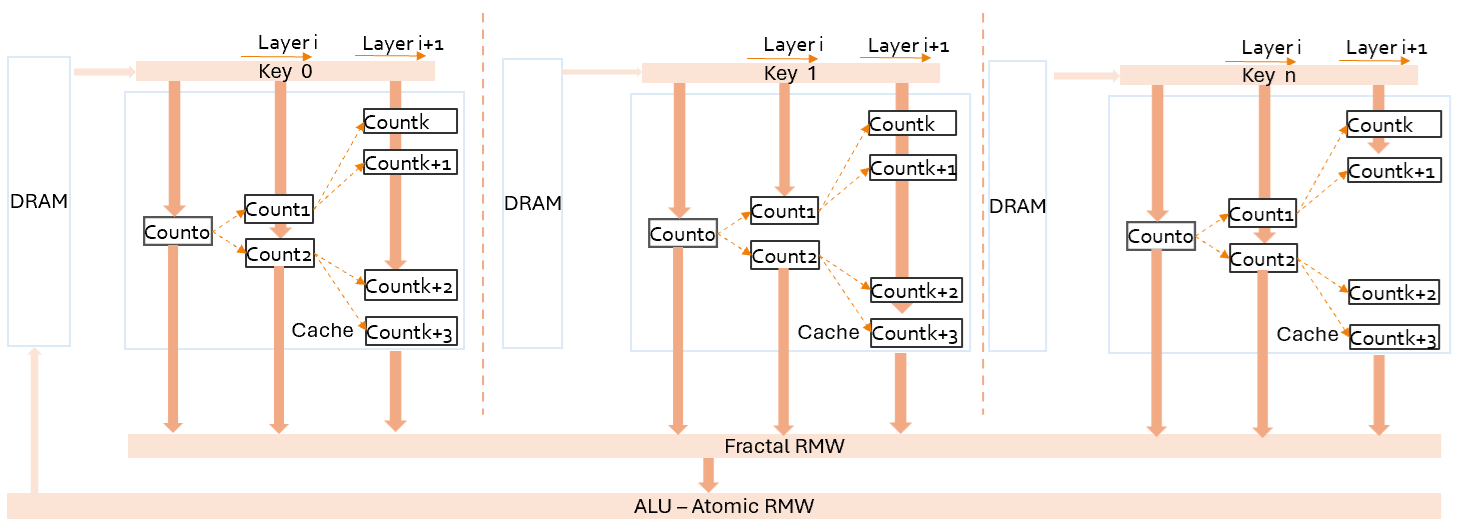}
    \caption{FractalSortCPU Logical Layout $p=16$}
    \label{fig:fractal_circuit}
\end{figure*}

Distribution dependence mitigation strategies such as sample sort and histogram sort incur pre-processing costs including increased memory footprint, which are mitigated by dependency graphs and stochastic sampling which predict load imbalance and perform smoothing techniques while relying on costly sampling processes \cite{Sanders1997} \cite{Dehne1996} \cite{Ajwani2016} \cite{Shi2012}. Such strategies increase memory footprint and pre-processing latency depending on the distribution, with significant variance between best case and worst case performance on skewed, uniform and partially sorted data \cite{DistribDepPerfMetricsDistribDepli2025}. \\

Limited parallelism due to multiple passes is mitigated by balanced buckets and increased bandwidth allocation where near-linear performance is achieved \cite{Bingmann2018}. Precision-scaling latency trade-off occurs in high-precision sorting scenarios which increase the number of passes to implement in-place cache-conscious algorithms where memory is limited due to histogram prefix arrays proportional to the radix size, permutation buffers for scatter stability and dependency graph data structures for distribution tracking \cite{hrs}. Increasing and reducing the per-pass radix size trades off the number of passes for memory usage, while data skew causes larger bucket size variance affecting memory and tail latency, mitigated by disaggregated radix-tree layouts which minimize I/O without fully resolving the trade-off \cite{radixTreeDM}.  \\

Partitioning, sampling and merging are used together to divide data sets into buckets for efficient sorting on limited compute resources used in sample sort and histogram sort, enabling balanced partitions and parallelization \cite{hrs}. $k$-way and cache-aware merging reduce passes but become the bottleneck as $k$ increases, and external memory or streaming benefit from larger sorted runs at the cost of large $k$, and extra passes trade-off bandwidth for peak memory, with sampling and partitioning enabling distribution agnosticism at the cost of memory, pre-processing and merge latency \cite{Bingmann2018}. \\

Radix sort is the state of the art for hardware-based FPGA and GPU sorting, and is affected by exponential histogram growth due to both data set size and precision, and the distribution dependence due to bucketing, making memory-bandwidth a significant limiting factor \cite{hrs}. The disaggregated radix tree helps reduce bandwidth usage by eliminating redundant I/O operations \cite{radixTreeDM}. FractalSort \cite{FractalSortFPGA} addresses histogram growth on FPGA through a compressed sparse histogram achieving 2.5x bandwidth efficiency over state-of-the-art FPGA sorting algorithms. In-place radix sorts and parallel counting schemes scale poorly with 64-bit and higher precisions due to the increasing number of passes needed to increase the precision of partially sorted keys and achieve a high-precision sort \cite{IntelIPP} \cite{Inoue2007}. \\

This work solves these limitations by using a compressed sparse histogram approach eliminating the need for costly sampling, pre-processing and input bucketing, while bounding memory usage to sublinear growth, and performing a fully parallel localized key update algorithm that minimizes data transfers, improves SIMD utilization, and ensures high bandwidth efficiency \cite{hrs}. \\

\section{System Architecture}
The FractalSortCPU algorithm consists of key-based threads performing sequential updates to each layer of the histogram. The keys are grouped in batches and processed in parallel on CPU using SIMD vectorization via Numba JIT to maximize throughput as shown in Figure \ref{fig:fractal_circuit} \cite{numbaReference}. \\

\subsection{Local Thread Histogram Update}
Each key is mapped to a thread which updates node counter variables along the path encoded by the key bit sequence. Collisions are mitigated using atomic operations supported by the Arithmetic Logic Unit (ALU) \cite{singhal2008atomic}. \\

\subsubsection{High Precision Histogram Update}
In high-precision sorting where the histogram is too large to fit in the CPU last-level cache, with $n$ nodes and counter variables encoding $n$ keys, the update algorithm involves two steps \cite{IntelI7Reference}. In the first step, a local merge on CPU where node updates are computed from key bits, and the updates are aggregated in $O(log(n))$ into a single local compressed tree. In the second step, the on-chip tree is transferred to memory using a parallel $O(log(n))$ merge operation based on a depth-first traversal from the local parent node. \\

\subsection{Global CPU Histogram Update}
The thread-based updates are merged into a single global histogram by threads co-resident on the CPU. The global merge tree is stored both in main memory and in the CPU last-level cache (LLC) \cite{IntelI7Reference}. Multi-thread updates are coordinated using Atomic Read-Modify-Write (RMW) and the Fractal RMW \cite{singhal2008atomic}. \\

\subsubsection{Fractal RMW}
For each key $k$ of $p$ bits, the histogram tree is updated by walking a root-to-leaf path determined by the key's bit sequence. At each level $l \in \{0, \ldots, L-1\}$ where $L = \min(p, \lceil\log_2(n)\rceil)$, the algorithm reads bit $l$ of the key to select the left ($0$) or right ($1$) child, and atomically increments the counter at the selected node. The full path update for one key touches exactly $L$ nodes, each requiring one Atomic RMW operation. \\

On x86 architecture, each atomic increment is executed in the integer ALU pipeline as a single transaction: read the counter from the cache line into a register, compute the new value, and write it back with a cache-line lock when necessary \cite{singhal2008atomic}. For histogram nodes in the CPU LLC, these operations are served from cache with no DRAM traffic. \\

Contention between threads drops exponentially with depth: at the root node all $n$ keys contend on a single counter, but at level $l$ the $n$ keys are distributed across $2^l$ nodes, giving an expected $\frac{n}{2^l}$ updates per node. At level $l \geq \lceil\log_2(n)\rceil$, the expected contention falls below one, making leaf-level updates effectively contention-free. This exponential decay is what enables the fully parallel update without locks or synchronization barriers beyond the atomic instruction itself. \\

\subsection{Batch Size Selection}
The batch size parameter determines the balance between locality and parallelism, where large batch sizes reduce synchronization frequency and relatively lower the cost of thread scheduling while increasing the risk of cache overflow and cache misses. Smaller batches have low risk of cache overflow hence they enable cache reuse at the cost of increased synchronization frequency when switching between batches. Optimal batch size is tuned empirically for the target device demonstrating latency improvements of over $10\%$ as shown in Figure \ref{fig:batch_size_latency_serial_29}. \\

\subsection{Batch Counter Update}
The cached histogram from a terminating batch is reused by the next batch to enable faster batch synchronization due to reduced memory transfer. This leads to a reduction in memory usage as the number of batches processed increases, as shown in Figure \ref{fig:batch_size_memory}. \\

\subsubsection{Counter Width Compression and Overflow}
Essential to overall performance is the compression scheme involving counter width tapering in Algorithm \ref{counter_update}. The principle exploits the reduction of significant bits on a well-balanced tree as levels increase from root to leaf, determined by the logarithm of the sub-tree size at each node $w_{c,l}=O(\lceil log(n_l)\rceil)$, which over large $n$ reduces by one at each level with leaves having approximately less than an average of two significant bits for uniformly distributed keys following Algorithm \ref{counter_update} and Algorithm \ref{alg:get_value}. \\

The local and global updates have complexity $O(log(n))$ as shown in Table \ref{tbl:complexity}. \\

\begin{algorithm}[t]
    \caption{Insertion with Counter Width Tapering}
    \label{counter_update}
\begin{algorithmic}[1]
\Require Root of histogram tree, path $p$, index $i$, $value$
\Ensure Counters updated along key path
\Function{AddItem}{$root$,\:$p$,\:$i$,\:$value$}
    \State $node \gets \texttt{root}$
    \If{$node$ is \texttt{null}}
        \State \Return $node$
    \EndIf
    \State $node.counter = node.counter+value[i]$
    \If{$p[i] \le 0\,$}
        \If{$node.left$ is \texttt{null}}
            \State $node.left = $ \Call{NodeOrTuple}{i,value[i]}
        \EndIf
        \State $child = node.left$
    \Else
        \If{$node.right$ is \texttt{null}}
            \State $node.right=$ \Call{NodeOrTuple}{i,value[i]}
        \EndIf
        \State $child = node.right$
    \EndIf
    \State \Return \Call{AddItem}{$child$,\:$p$,\:$i+1$,\:$value$}
\EndFunction
\end{algorithmic}
\end{algorithm}

\subsection{Histogram Read and Write Operations}
Looking up the value at an index involves depth-first path traversal of the histogram from root to leaf node. Each node has a range of indices representing the leaves in its subtree. The child selected at each node is the one containing the desired index. Traversal stops when a matching leaf node is reached. The $getValue(index)$ hash function has $O(min(p, log(n)))$ bounded latency as shown in Algorithm \ref{alg:get_value}. \\ 

\begin{algorithm}[h]
\caption{Retrieving Items from Histogram}
\label{alg:get_value}
\begin{algorithmic}[1]
\Require Root of histogram tree, $index$, $defaultVal$
\Ensure Value at target index or $defaultVal$ if not found
\Function{GetItem}{$root$,\:$index$,\:$defaultVal$}
    \State $node \gets \texttt{root}$
    \If{$node$ is \texttt{null}}
        \State \Return $defaultVal$
    \EndIf
    \If{$\,node.counter \le index\,$}
        \State $child, val = node.left, node.value$
        \State \Return  \Call{GetItem}{$child$,\:$index$,\:$val$}
    \Else
        \State $child, val = node.right, node.value$
        \State $index = index-node.counter$
        \State \Return  \Call{GetItem}{$child$,\:$index$,\:$val$}
    \EndIf
\EndFunction
\end{algorithmic}
\end{algorithm}

Determining the index of a value involves the following $O(p)$ algorithm shown in Algorithm \ref{alg:get_index}, a significant improvement on the $O(log(n))$ search latency of array data structures. \\

\begin{algorithm}[h]
\caption{Retrieving Indices from Histogram}
\label{alg:get_index}
\begin{algorithmic}[1]
\Require Root of histogram tree, $value$
\Ensure Index of target value
\Function{GetIndex}{$root$,\:$value$,\:$defaultIdx$}
    \State $node \gets \texttt{root}$
    \If{$node$ is \texttt{null}}
        \State \Return $defaultIdx$
    \EndIf
    \If{$\,node.cnt \le targetIndex\,$}
        \State $child, idx = node.left, defaultIdx$
    \Else
        \State $child, idx = node.right, defaultIdx+node.cnt$
    \EndIf
    \State \Return  \Call{GetIndex}{$child$,\:$value$,\:$idx$}
\EndFunction
\end{algorithmic}
\end{algorithm}

\subsection{Bandwidth Efficiency}


\subsubsection{Computable Pointer Optimization}
The histogram nodes are optimized for bandwidth usage by employing a format that stores nodes without identifier information, achieved by defining a subset of histogram nodes at configurable maximum depth $L_c=32$ with identifiers determined by memory location, eliminating the need for explicit identifiers of size $w_i=O(\frac{p}{\lceil log(n)\rceil})$ and pointers of size $w_r=O(log(n))$, enabling the sparse section of the histogram to be updated bandwidth-efficiently using only counter variables of size $w_{c,l}=O(\lceil log(n)-l\rceil)$ for level $l \in \{0, log(n)\}$ as shown in Algorithm \ref{alg:new_node_or_tuple}.  \\

\begin{algorithm}[H]
\caption{New Node Creation and Resolution}
\label{alg:new_node_or_tuple}
\begin{algorithmic}[1]
\Require Level and counter value $l$, $counter$
\Function{NodeOrTuple}{$l$,\:$counter$}
    \State $node \gets \texttt{root}$
    \If{$l \leq L_c$}
        \Comment{Write to computable location}
        \State \Return $\{\textbf{counter}: counter\}$
    \Else
        \Comment{Allocate a new node and return its pointer}
        \State \Return  \Call{NewNode}{$counter$}
    \EndIf
\EndFunction
\end{algorithmic}
\end{algorithm}

\subsection{FractalSortCPU Complexity}

\subsubsection{Complexity}

The sort complexity where $p$ is the precision, $n$ is the number of inputs in a batch, and $n_L=\frac{W}{w_c}$ is the number of $W$-bit registers needed to store $p$ counters of $w_c \approx \frac{log(n)}{2}$ size along the key path, is approximately $O(n\lceil\frac{p}{n_L}\rceil)$ serial and $O(2min(\lceil\frac{p}{n_L}\rceil,2log(n)))=O(2min(\lceil\frac{pw_c}{W}\rceil, 2log(n)))$ complexity since each key has to update its associated path in the histogram, which represents either a sparse trie with depth $L_n=log(n)$ or full histogram with depth $L=p$ as shown in Table \ref{tbl:complexity}. \\

\subsubsection{Memory usage}
The storage space utilized by the algorithm is determined by the size of the sparse histogram \\ 
\[
O(
\left\{
    \begin{array}{lr}
        nw_c, & \text{if } n \leq 2^{p} \\
        2^{p}w_c, & \text{if } n > 2^{p} \\
    \end{array}
\right\})
\] \\

\begin{table}[h]
\centering
\begin{tabular}{ |p{1.01cm}|p{1.1cm}|p{0.69cm}|p{1.6cm}|p{2.590cm}|  }
 \hline
 \multicolumn{5}{|c|}{ Complexity Comparison } \\
 \hline\hline
 HRS  & GPU Terasort \& Bitonic Network & Paradis & Bonsai & FractalSortCPU \\
 \hline
$O(\lceil \frac{k}{d} \rceil  n)$ & $O(log^{2}n)$ & $O(\frac{n}{P})$ & $O(n  \lceil log_{l}(n) \rceil)$ & $O( 2 min(p, 2 log(n)))$  \\
 \hline
\end{tabular}
\caption{FractalSortCPU Complexity Summary \cite{hrs} \cite{paradis} \cite{Bonsai}}
\label{tbl:complexity}
 \end{table}

\subsubsection{FractalSortCPUA Algorithm}
The FractalSortCPUA algorithm produces a sorted array of keys from the histogram in a bandwidth-efficient streaming operation shown in Algorithm \ref{alg:reconstruct}. The algorithm iterates over all $2^{l_n - l_b}$ bins in tree-walk order for trie depth $l_n=L_n$ and bin width $l_b=log(n_{bins})$. The last level of the histogram contains the counter variable associated with each bin $C$. For each bin $b$, the bin identifier bits are recovered by reversing the tree ordering of the bin index. Within each bin, the sorted entries are traversed using the index array which maps each sorted position to an arrival position. Each entry is decoded by extracting the trailing bits and tree-reversed offset, and the full $p$-bit key is reconstructed by combining the bin identifier bits, offset bits and trailing bits. \\

\begin{algorithm}[t]
    \caption{FractalSortCPUA: Sorted Array Reconstruction}
    \label{alg:reconstruct}
\begin{algorithmic}[1]
\Require Entries $E$, index array $I$, bin counts $C$, precision $p$, tree depth $l_n$, bin depth $l_b$
\Ensure Sorted array $K$ of $n$ reconstructed $p$-bit keys
\Function{ReconstructAll}{$E, I, C, p, l_n, l_b$}
    \State $n_{bins} \gets 2^{l_n - l_b}$, $t \gets p - l_n$
    \State $offset \gets 0$, $idx \gets 0$
    \For{$b = 0$ \textbf{to} $n_{bins} - 1$}
        \If{$C[b] = 0$} \textbf{continue} \EndIf
        \State $key_{low} \gets$ \Call{BitReverse}{$b,\: l_n - l_b$}
        \For{$s = 0$ \textbf{to} $C[b] - 1$}
            \State $entry \gets E[offset + I[offset + s]]$
            \State $trailing \gets entry \;\&\; (2^{t} - 1)$
            \State $tree_{off} \gets entry \gg t$
            \State $off_{val} \gets$ \Call{BitReverse}{$tree_{off},\: l_b$}
            \State $K[idx] \gets key_{low} \;|\; (off_{val} \ll (l_n - l_b)) \;|\; (trailing \ll l_n)$
            \State $idx \gets idx + 1$
        \EndFor
        \State $offset \gets offset + C[b]$
    \EndFor
    \State \Return $K$
\EndFunction
\end{algorithmic}
\end{algorithm}

The outer loop is $O(n_{bins})$ and the inner loop process $n$ entries exactly once with $O(1)$ work per entry (index lookup, bit extraction, bit-reverse via LUT, and concatenation), giving total complexity $O(n)$. Memory traffic is one sequential read of the index array and one indexed read of the entry array, totaling $\approx 2 \times \frac{p}{8}$ bytes per key. The FractalSortCPUA algorithm demonstrates $41\%$ bandwidth efficiency as shown in Figure \ref{fig:bandwidth_efficiency}. \\

\section{Evaluation}
The experiments are performed on a Windows 11 Pro device with an Intel(R) Core(TM) i7-8665U CPU at 1.90GHz capable of 2.11GHz processing with 32 GB RAM, 950 GB SK Hynix SSD storage and Intel Emerald Rapids 32 vCPU Debian 120GB RAM host for benchmarking \cite{IntelI7Reference}. The runtime environment includes 64 bit Python 3.9.1, and the tests involve comparison of FractalSortCPU to state-of-the-art CPU-based sorting algorithms baseline implemented in Numba: quick sort, merge sort, heap sort, Tim sort, and a custom in-place implementation of radix sort with each reported result a statistical average of 10 runs \cite{numpyReference}. Analysis of the effect of tuning batch size on latency and memory usage is performed on different data set sizes. Comparison of FractalSortCPU to the state of the art involves data sets of $n\in \{2^{10} \to 2^{31}\}$. \\

\subsection{FractalSortCPU Experiment Test Bed} 
\begin{figure}[h]
    \centering
    \includegraphics[scale=0.885]{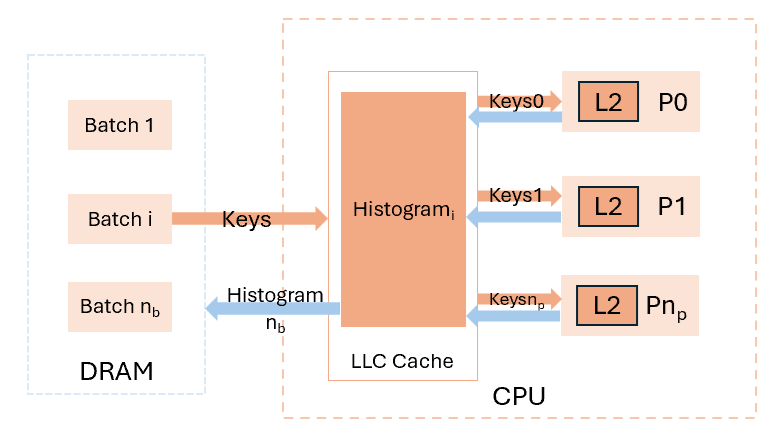}
    \caption{FractalSortCPU Multi-batch Test Bed $p=16$}
    \label{fig:experimental_setup}
\end{figure}

The experiment setup includes a random number generator which populates batches of keys loaded to CPU and processed as shown in Figure \ref{fig:experimental_setup}. \\

Although the keys are uniformly distributed, the histogram tree structure provides some robustness to skew since each key independently walks its own root-to-leaf path with no input bucketing. However, the counter width tapering scheme assumes approximately balanced subtrees where counter widths decrease as $w_{c,l} = O(\lceil\log(n) - l\rceil)$. Under heavy skew (e.g., Zipfian), subtree sizes may deviate from this assumption, and empirical evaluation across non-uniform distributions is left to future work. \\

\subsection{Latency Analysis}
\begin{figure*}[!htbp]
    \centering
    \includegraphics[scale=0.5155]{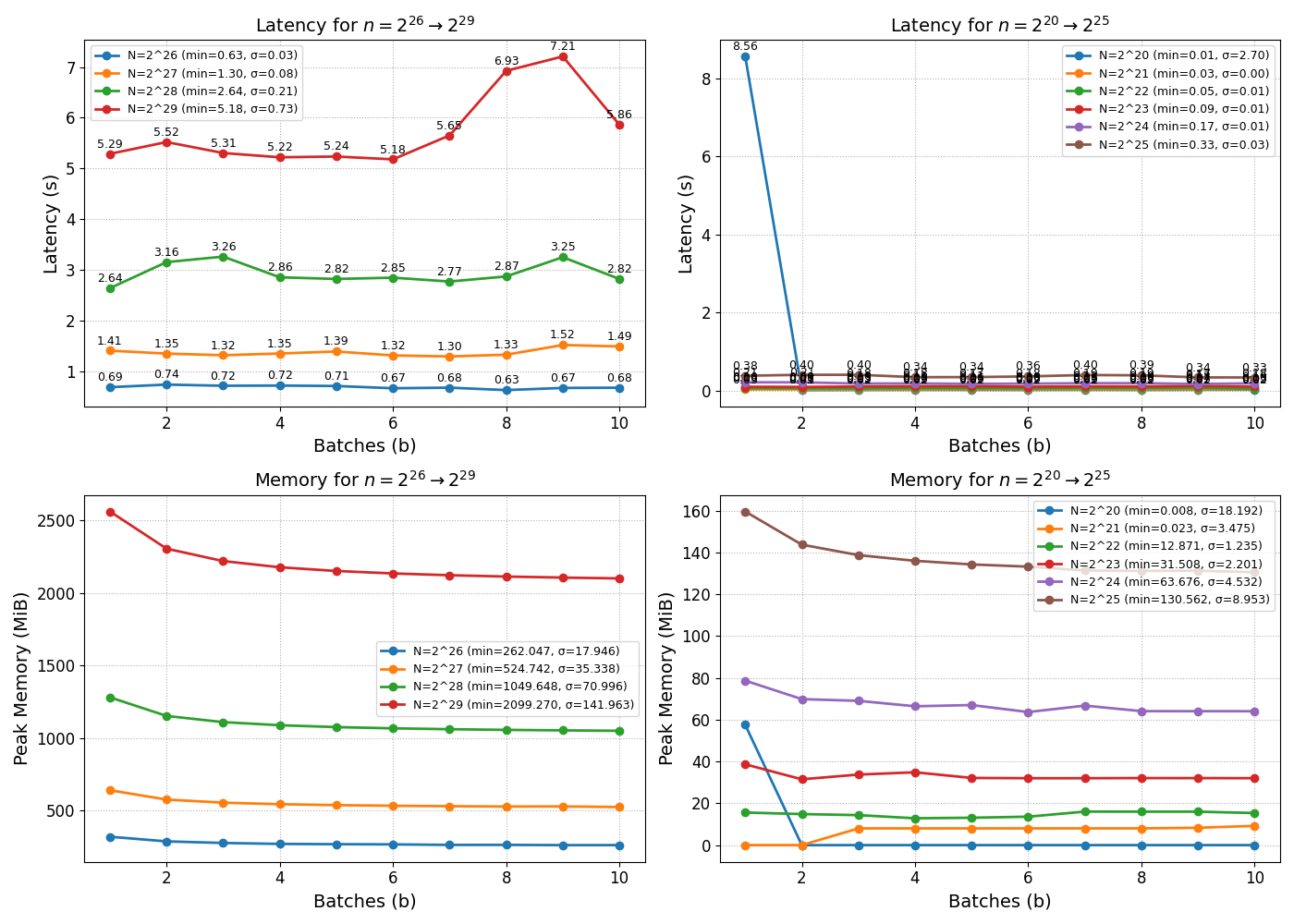}
    \caption{FractalSortCPU Latency and Memory for  Number of Parallel Batches $b \in 1 \to 10$}
    \label{fig:batch_size_latency}
\end{figure*}

The latency curve for FractalSortCPU has sub-linear growth compared to the baselines and is $34\%$ faster than the next best radix sort at $n=2^{29}$ as shown in Figure \ref{fig:batch_size_latency} and Table \ref{table:sort_latency}. FractalSortCPU demonstrates lowest latency at $n\geq 2^{21}$. This speed-up is due to the reduced number of radix passes on compressed entries, with the bandwidth savings from histogram compression becoming apparent for large $n$. \\

\begin{figure*}[!htbp]
    \centering
    \includegraphics[scale=0.5155]{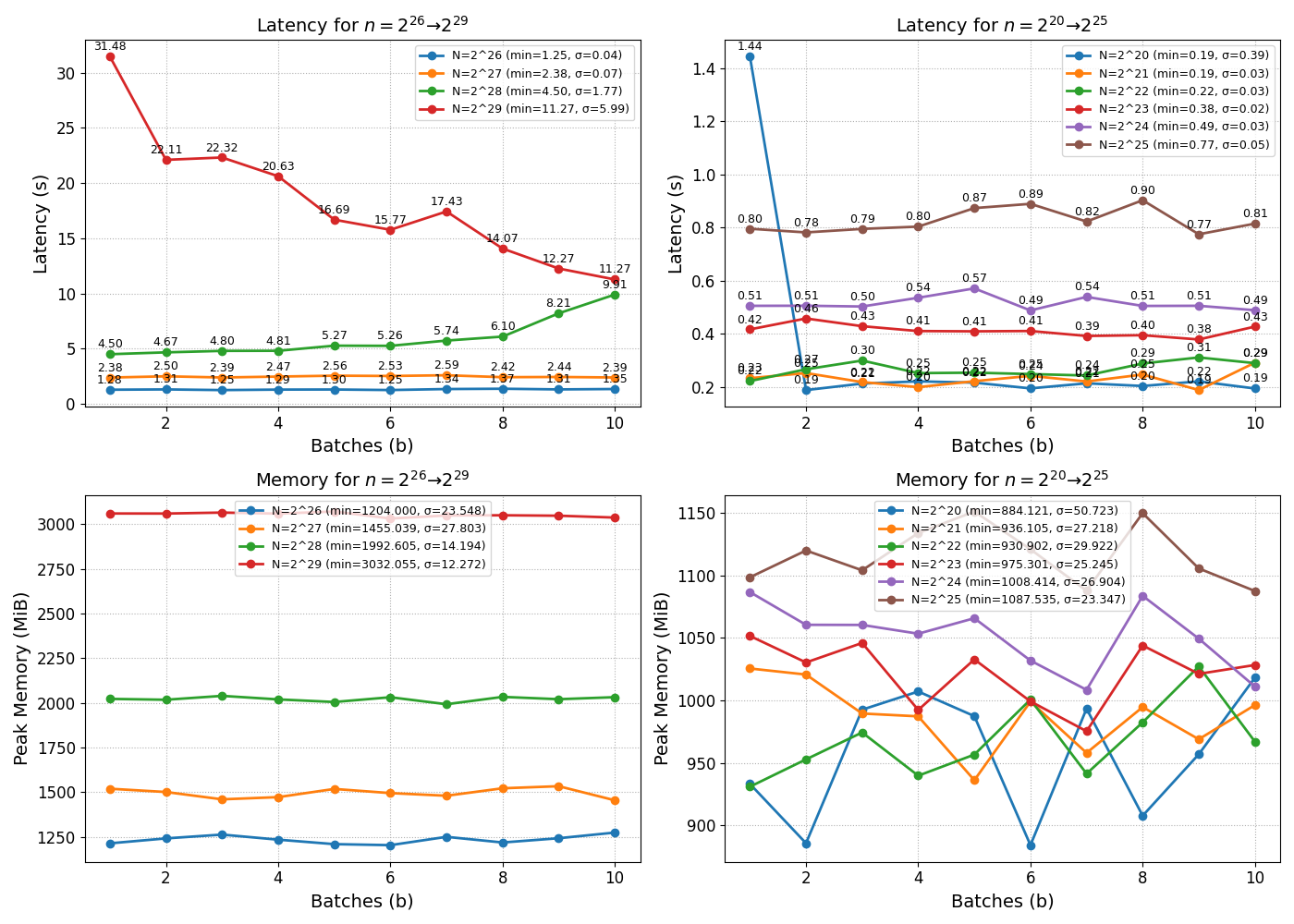}
    \caption{FractalSortCPU Latency and Memory for  Number of Serial Batches $b \in 1 \to 10$}
    \label{fig:batch_size_latency_serial}
\end{figure*}

\begin{table}[h]
\begin{tabular}{| p{1.78cm}| p{0.69cm}| p{0.69cm}| p{0.65cm}| p{0.72cm}| p{0.68cm}| p{0.68cm}|}
\hline

Algorithm & Quick sort & Merge sort & Heap sort  & Radix sort & Fractal sort \\
 \hline\hline
Latency (s) &	59.51 & 80.36 & ---
 & 8.38 & 5.53 \\
\hline
Memory (MiB) &	$4.10\times 10^3$ & $4.10\times 10^3$ & --- &	$6.14\times 10^3$ &	$4.61\times 10^3$ \\
\hline
\end{tabular}
\caption{Sort latency (seconds) and Memory Usage (MiB) for $n=2^{29},p=32$ on CPU with Numba}
\label{table:sort_latency}

\end{table}

\begin{figure}[H]
    \centering
    \includegraphics[scale=0.3315]{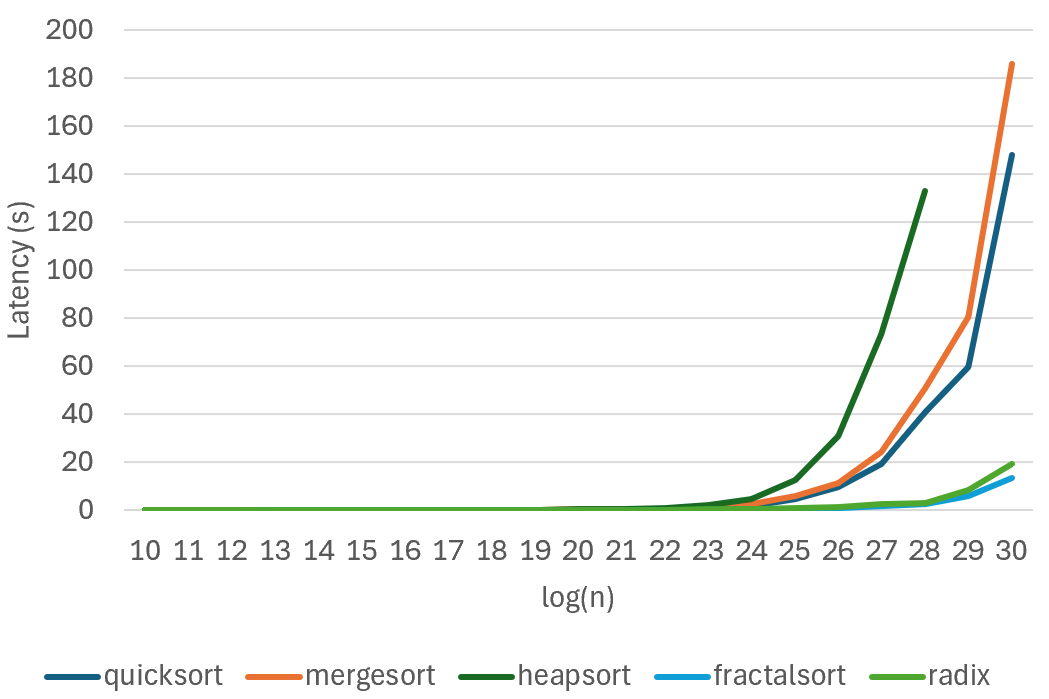}
    \caption{FractalSortCPU Latency for $n=2^{30},p=32$}
    \label{fig:latency_1gb}
\end{figure}

\subsection{Memory Usage Analysis}
The memory usage analysis shows a 7x higher footprint for FractalSortCPU for small data sets which remains approximately constant for $n\leq 2^{15}$, at which point radix sort has higher utilization. This is because the histogram size is dependent on $log(n)$, and due to the histogram cache between batch updates creating a trade-off between memory and latency as shown in Figure \ref{fig:memory_1gb}. At $n=2^{29}$ FractalSort consumes 4,608 MiB ranking second after the in-place comparison sorts (keys $4n$ + entry array $4n$ + radix temp buffers $\approx n$ bytes), and $25\%$ less than radix sort at 6,144 MiB. At $n=2^{30}$ FractalSort uses 9,216 MiB, maintaining the $25\%$ memory advantage over radix sort as $n$ increases. \\

\subsection{Batch Size Optimization}
Streaming the dataset in input order in batches of equal size for more efficient processing with respect to the sorting resources available demonstrates varying benefits regarding latency and memory utilization depending on how the batches are processed both serially and in parallel as shown in Figure \ref{fig:batch_size_latency} and Figure \ref{fig:batch_size_latency_serial}. \\

\subsubsection{Parallel Batch Processing}
Processing batches in parallel involves dedicating a thread to each batch on the limited CPU thread pool. Increasing the number of batches has minimal impact on latency and memory usage for most data sets with latency standard deviation $\sigma \in 0.03s \to 0.38s \approx 6\% \to 12\%$ and memory standard deviation $\sigma \in 3\times 10^{-6} \to 14\%$. \\

\begin{figure}[H]
    \centering
    \includegraphics[scale=0.375]{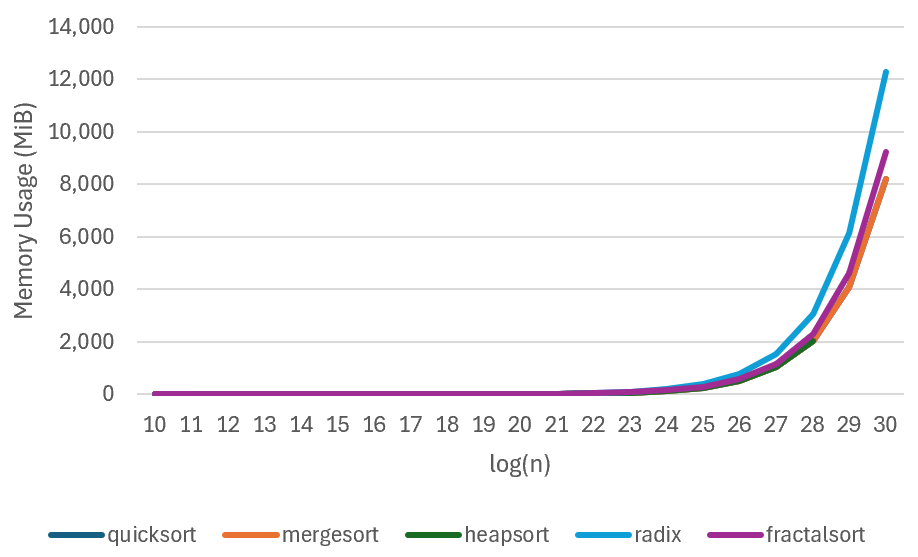}
    \caption{FractalSortCPU Memory Usage for $n=2^{10} \to 2^{30}$ at $p=32$}
    \label{fig:memory_1gb}
\end{figure}

An exception is $n=2^{25}$ and $n=2^{20}$ which show significant $20\%$ reduction in latency at $b=6$ and $20KB$ reduction in memory $b=2$ respectively, as shown in Figure \ref{fig:batch_size_latency}. \\

\subsubsection{Serial Batch Processing}
\begin{figure}[H]
    \centering
    \includegraphics[scale=0.38155]{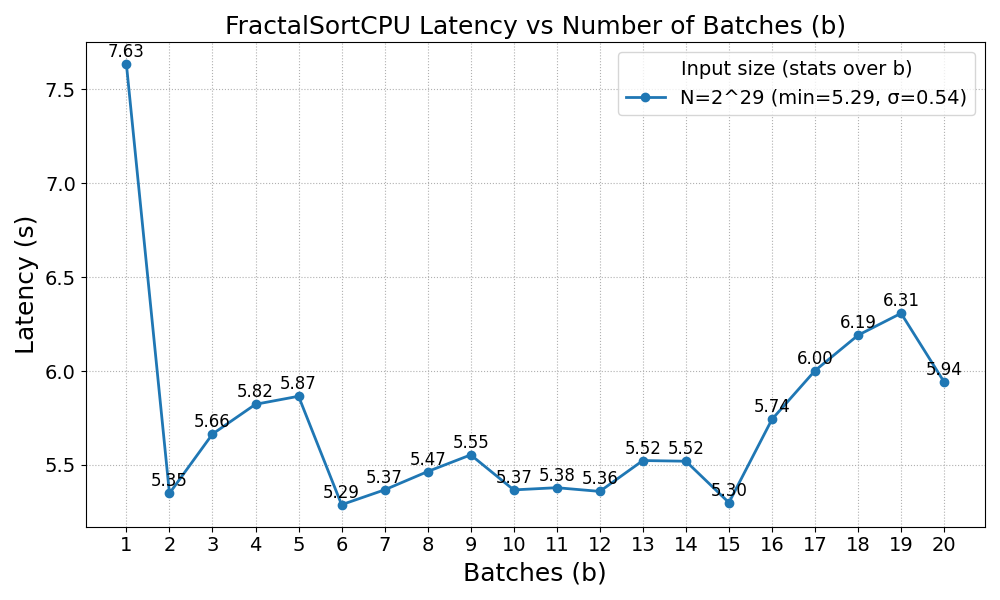}
    \caption{FractalSortCPU Latency for $n=2^{29}$ and Serial Batch $b \in 1 \to 20$}
    \label{fig:batch_size_latency_serial_29}
\end{figure}

\begin{figure}[H]
    \centering
    \includegraphics[scale=0.365]{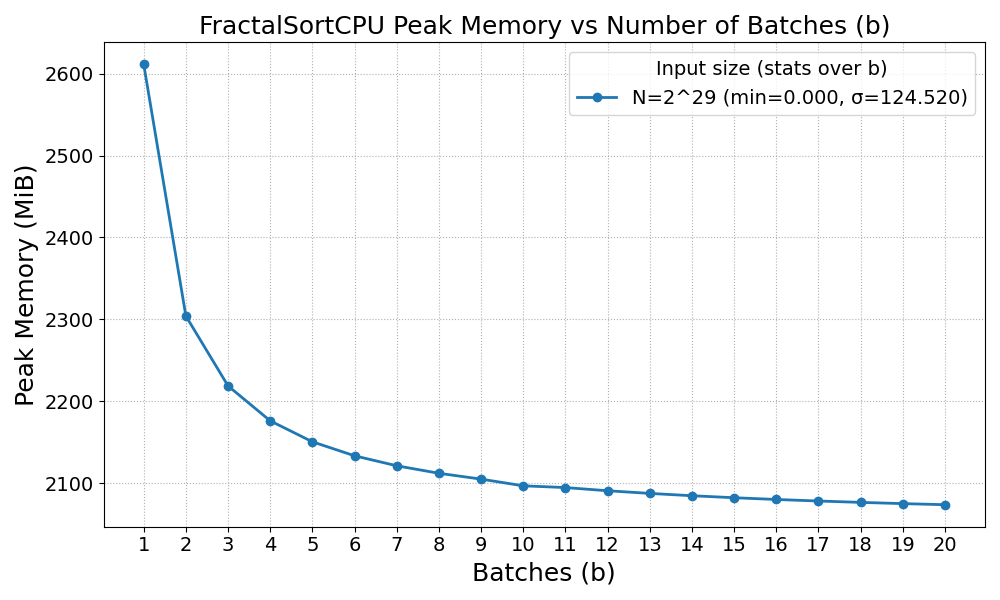}
    \caption{FractalSortCPU Memory Usage for $n=2^{29}$ and Serial Batch $b \in 1 \to 20$}
    \label{fig:batch_size_memory}
\end{figure}

Processing batches in serial is simpler with no thread creation overhead leading to lower latency for most data sets. Increasing the number of batches has moderate impact on latency for $b \geq 2$, with latency standard deviation $\sigma = 0.54s$ across $b \in 1 \to 20$ at $n=2^{29}$. Latency at $b=1$ is higher due to larger per-bin radix sort temporary buffers, dropping from $7.63s$ to $5.35s$ at $b=2$ for all $n\neq 2^{28}$, averaging $\approx 5.6s$ for $b \geq 2$ with periodic variation as shown in Figure \ref{fig:batch_size_latency_serial_29}. Memory decreases monotonically from 2,612 MiB at $b=1$ to 2,074 MiB at $b=20$, a $20\%$ reduction, with diminishing returns beyond $b=10$ as shown in Figure \ref{fig:batch_size_memory}. \\

\begin{figure}[H]
    \centering
    \includegraphics[scale=0.7085]{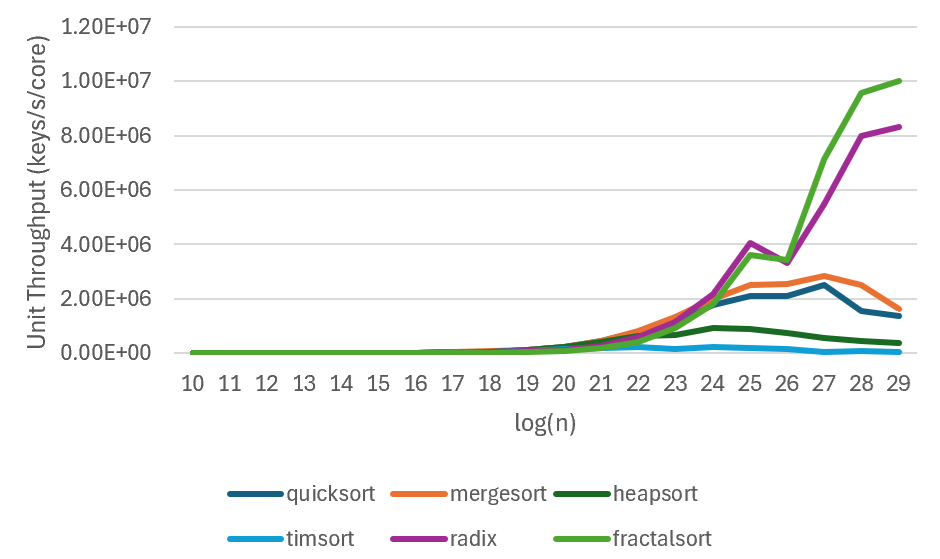}
    \caption{Unit Throughput for $2^{10}\leq n \leq 2^{30}$}
    \label{fig:unit_throughput}
\end{figure}

\begin{figure}[H]
    \centering
    \includegraphics[scale=0.715]{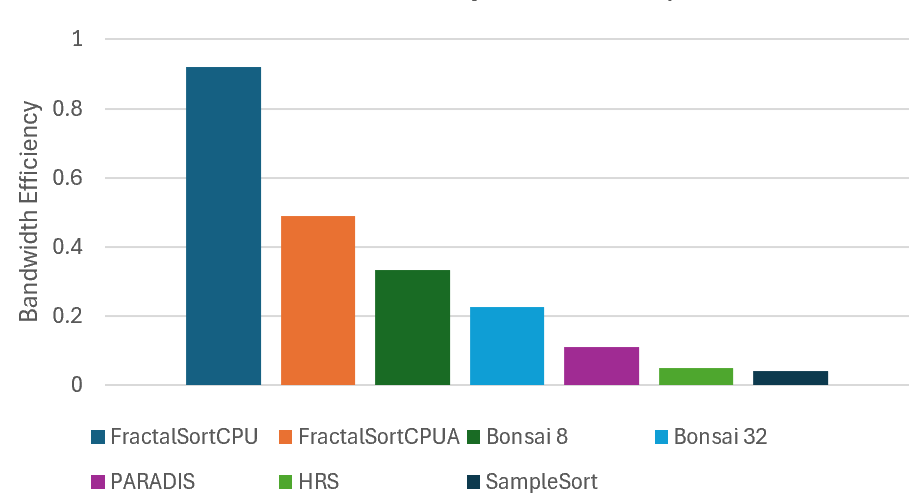}
    \caption{Bandwidth Efficiency for 16GB datasets ($p=16$) compared to Bonsai \cite{Bonsai}, Paradis \cite{paradis}, HRS \cite{hrs} and SampleSort \cite{samplesort}}
    \label{fig:bandwidth_efficiency}
\end{figure}

\subsection{Unit Throughput}
The unit throughput is calculated as $T_{unit}=\frac{n}{t\times n_{c}}$ where $t$ is the latency and $n_{c}$ is the number of CPU cores (4 cores on 8 threads). FractalSortCPU has the highest unit throughput at $n\geq 2^{26}$ growing to $1\times 10^{7}$ keys/s, a $20\%$ speed up over the second best radix sort at $n=2^{30}$ at $p=32$ as shown in Figure \ref{fig:unit_throughput}. State-of-the-art CPU-based Paradis \cite{paradis} sorts $4GB$ of $p=16$ data in $4.6s$ on a 32-core Intel Xeon (E7-8837) processor for unit throughput $14.59\times 10^{6}$ keys/s, while FractalSortCPU latency is $21.22s$ at $b=14$ for $25.30\times10^6$ keys/s, a $73.4\%$ speed up.






\subsection{FractalSortCPU on Large Datasets}
For large data sets $n\geq2^{30}$ FractalSortCPU exhibits low latency $t_{2^{30}}=13.4$s in contrast to the comparison sorts which demonstrate superlinear growth as shown in Figure \ref{fig:latency_1gb}. Memory usage increases linearly with $n$ at $9n$ bytes, $25\%$ lower than radix sort at $12n$ bytes as shown in Figure \ref{fig:memory_1gb}. The better performance of FractalSortCPU at large $n$ is due to its latency and memory usage being primarily determined by histogram size which grows sublinearly in $n$ as $O(2^p\log_2(n))$, determined by the counter variable width $\approx \log_2(n)$ stored in each node as shown in Figure \ref{fig:memory_1gb}. \\

\subsubsection{Bandwidth Efficiency}
Following the bandwidth-efficiency analysis used in prior sorting work \cite{hrs} \cite{FractalSortFPGA}, we define bandwidth efficiency as the ratio of actual throughput to total memory (DRAM) bandwidth consumed:
\begin{equation}
b_{eff} = \frac{T_{actual}}{B_{DRAM}}
\label{eq:beff}
\end{equation}
where $T_{actual}$ is the sum of read and write throughput in bytes per second for $n$ keys of $p$ bits sorted in time $t$, and $B_{DRAM}$ is the total DRAM traffic in bytes per second including all intermediate reads and writes. The ratio is dimensionless with higher values indicating more efficient use of memory bandwidth. Algorithms that require multiple passes over the data or store large intermediate structures have lower $b_{eff}$ due to increased $B_{DRAM}$. \\

FractalSortCPU is compared to state-of-the-art algorithms over a variety of data set sizes to explore on-chip and off-chip memory access and sort performance. The bandwidth efficiency of FractalSortCPU is significantly higher than the state of the art due to the small size of the compressed histogram which fits entirely on CPU LLC cache at $p=16$. Compared to state-of-the-art sorting algorithms, FractalSortCPU performs very favorably with a 6x increase in bandwidth efficiency compared to CPU \cite{paradis}, 3x compared to GPU \cite{hrs} and 2.5x compared to FPGA \cite{Bonsai} for 16GB data sets as shown in Figure \ref{fig:bandwidth_efficiency}. \\

\section{Conclusion and Recommendations}
FractalSortCPU introduces a histogram compression scheme demonstrating low latency and sublinear memory growth as dataset size increases. It performs better for large $n$, hence suited for large dataset processing where latency and memory usage are key concerns. Though in-place radix sort shows superior performance for medium datasets and comparison-based sorters for small datasets, FractalSortCPU scales more slowly for large datasets with $25\%$ lower memory usage compared to radix sort at $n=2^{29}$. \\

The key contributions of FractalSortCPU include a space-efficient fractal sorting algorithm, involving a precision-depth bound histogram tree of sparse counters, and a parallel merge operation that eliminates the need for data pre-processing and input partitioning. Caching of the histogram allows the sharing of merge tree between batches enabling the amortization of batch-switching costs as the number of batches increases with a $20\%$ reduction in batch memory usage at $b=20$. \\ 

FractalSortCPU provides significant improvement over state-of-the-art sorting algorithms on CPU, GPU, and FPGA by 6x, 3x, and 2.5x in bandwidth efficiency on 512MB to 32GB data sets, a key benefit of using this algorithm, and demonstrates a $73.4\%$ improvement in unit throughput compared to the state-of-the-art CPU algorithm Paradis as shown in Figure \ref{fig:unit_throughput}. This is because no input partitioning is needed which eliminates round trips to memory storage required during pre-processing, achieving a 3x improvement in bandwidth efficiency over state-of-the-art Bonsai on 16KB-32GB data sets as shown in Figure \ref{fig:bandwidth_efficiency}. Optimizing the count variable width $w_{c}$ based on the reduction of subtree size with increasing merge tree depth improves bandwidth efficiency and data compression.  \\


FractalSortCPU is useful for sorting data streaming scenarios, and implementation on CPU enables both significant performance improvements and greater accessibility due to lower adoption costs. Together with the FPGA implementation \cite{FractalSortFPGA}, the FractalSort family provides bandwidth-efficient sorting across heterogeneous platforms. Future work will involve exploration of the GPU platform for higher parallelism and data access rates and exploration of novel high-precision sorting optimizations.




{\large

}

\end{document}